\documentclass[preprint,authoryear,12pt]{elsarticle}

\usepackage[english]{babel}
\usepackage[latin1]{inputenc}
\usepackage[T1]{fontenc}
\usepackage{amsfonts}
\usepackage{amsmath}
\usepackage{mathrsfs}
\usepackage{latexsym}
\usepackage{natbib}
\usepackage{epsfig}

\newcommand{\mytitle}{Optimal forest rotation age under efficient climate change mitigation}

\begin{document}

\begin{center}
	\huge \mytitle
	\normalsize

  \vspace{0.5cm}
	Tommi Ekholm\footnote{Email address: tommi.ekholm@vtt.fi, Tel.: +358 40 775 4079}
	
  VTT Technical Research Centre of Finland, P.O.Box 1000, FIN-02044 VTT

  \vspace{0.5cm}
	\small
	October 28$^{th}$ 2015
  \vspace{0.25cm}

	Preprint of an article in Forest Policy and Economics, \\
	doi: 10.1016/j.forpol.2015.10.007


\end{center}


\section*{Abstract}

\leftskip=1cm
\rightskip=1cm

This paper considers the optimal rotation of forests when the carbon flows from forest growth and harvest are priced with an increasing price. Such evolution of carbon price is generally associated with economically efficient climate change mitigation, and would provide incentives for the land-owner for enhanced carbon sequestration. For an infinitely long sequence of even-aged forest rotations, the optimal harvest age changes with subsequent rotations due to the changing carbon price. 
The first-order optimality conditions therefore also involve an infinite chain of lengths for consecutive forest rotations, and allow the approximation of the infinite-time problem with a truncated series of forest rotations.

Illustrative numerical calculations show that when starting from bare land, the initial carbon price and its growth rate both primarily increase the length of the first rotation. 
With some combinations of the carbon pricing parameters, the optimal harvest age can be several hundred years if the forest carbon is released to the atmosphere upon harvest. 
In the near term, however, a higher growth rate of carbon price can lead to shorter rotations for forests that are already near their optimal rotation age, indicating that the effect of carbon price dynamics on optimal rotation is not entirely monotonous.
The introduction of carbon pricing can also have a significant impact on bare land value, and in some considered parametrizations the land value was based solely on its potential to capture and store atmospheric carbon.

\leftskip=0pt\rightskip=0pt



\section{Introduction}

Forests are in a natural interaction with atmospheric carbon dioxide, the main driver behind anthropogenic climate change. A growing tree stores carbon from the air in itself, which is later released back to the air through fires, natural decay of the biomass, or human-induced activities. Forests involve globally both large stocks and flows of carbon, making them an important element in the context of climate change and its mitigation.

Achieving stringent targets of climate change mitigation -- e.g. the 2°C limit considered currently in the United Nations' climate negotiations -- have been estimated to require significant economic effort. Therefore it is of importance that mitigation is implemented in an economically efficient manner, putting the mitigation resources in the best possible use. A general view in economics states that this could be achieved by pricing all carbon dioxide flows to and from the atmosphere with a uniform price across the economy.

Such pricing can be implemented in forestry by crediting a forest owner for the carbon absorbed due to forest growth, and taxing for the carbon that is released back to the atmosphere. The latter would occur e.g. if the tree biomass is harvested and combusted for energy, through the gradual decay of forest products and litter, or due to forest fires. On the other hand, if the harvested wood's carbon is stored permanently, the atmospheric release and the subsequent levy for the forest owner could be avoided -- perhaps at least partially -- creating an incentive for the long-term storage of the harvested wood's carbon.

Pricing of forest carbon flows and its implications for forest management have been studied previously in forest economics. 
The optimal rotation age of even-aged stands under constant carbon pricing has been examined e.g. by \citet{Plantinga1994}, \citet{vanKooten1995} and \citet{Hoen1997}.
Later contributions have also considered additional forest management options, such as fertilization \citep{Stainback2002} and thinnings \citep{Pohjola2007}. 
Analyses of forestry carbon pricing as a part of the larger macroeconomy have also been presented \citep{Tahvonen1995, Sohngen2003}.
A main conclusion from these studies has been that the pricing of forest carbon lengthens the rotations on the stand-level, leading to a larger forest carbon stock and thus a higher amount of sequestered atmospheric carbon.

The stand-level analyses referenced above have assumed a constant price for carbon.
However, an economically efficient climate policy generally implies an increasing carbon price. This is a common result from numerous numerical scenarios addressing efficient climate change mitigation \citep[see e.g.][]{Nordhaus2010}. An analytical solution for limiting the temperature increase below 2°C in a cost-efficient manner suggested that the carbon price increase should be close to exponential for several decades \citep{Ekholm2014}.
Although an increasing carbon price has been a part of e.g. macroeconomic approaches using intertemporal optimization \citep{Sohngen2003} and an analysis on the decision to convert land to carbon storage through afforestation \citep{vantVeld2005}, 
this feature has been absent from stand-level analyses of optimal forest rotation length.

The purpose of this paper is to analyse the forest-owner's problem of maximizing the net present value of forest revenues from even-aged rotations in a case where the changes in tree carbon stock are priced with an exponentially increasing price.
The paper provides a generalization to the constant-price approach of \citet{vanKooten1995}, employing the same problem formulation 
and numerical parameters to allow a direct comparison with their results.
With the consideration of exponentially increasing prices -- contrasting from the case of constant prices -- the problem setting described here is not stationary, but changes over time as the carbon price grows. This implies that the optimal harvest ages differ on subsequent rotations, and the optimal rotation length for the same type of forest changes over time.

The paper is structured so that the problem setting is described in section \ref{sec:analytical}, followed by a derivation for the first-order optimality conditions. Section \ref{sec:numerical} provides numerical examples of optimal rotation ages when starting from bare land, the implications of increasing carbon pricing to bare land value, and the currently optimal harvest ages; using the forest growth curves and parameters from \citet{vanKooten1995}.
The role of these calculations is to illustrate the problem setting -- how carbon pricing under economically efficient climate change mitigation might affect forest economics -- and further research should analyse optimal strategies for various actual forest stands.
Finally, section \ref{sec:conclusions} discusses the results' implications on a broader context.

\section{An analytical consideration}
\label{sec:analytical}

The considered forest-owner's optimization problem is to maximize the present value of net revenues from even-aged rotations of a forest plot that is initially bare land, when timber price remains constant and all carbon flows to and from the atmosphere are priced with an exponentially increasing price.
Each harvest yields a volume of wood that can be sold at a constant price, but also necessitates a costly replanting of the subsequent forest stock.
The forest-owner is credited for each tonne of carbon sequestered by the forest during its growth, and is taxed for each tonne of carbon released back to the atmosphere, both with a carbon price that increases exponentially. The release is assumed either to take place immediately after harvest, disregarding temporary stocks of forest products or carbon in the soil; though allowing that a certain fraction of the harvested carbon will be stored indefinitely and thus not being levied for the atmospheric carbon release.
Knowing the assumed growth of the carbon price with perfect foresight, the forest-owner chooses an infinite sequence of rotation lengths $T_1$,  $T_2$, $T_3$... that maximize the value of the bare land.

The maximization problem, defined at time $t$, can be formulated as:
\begin{eqnarray}
\label{eq:maxproblem}
\emph{max}_{\,T_1,T_2,...} && \sum_{i=1}^\infty  \bigg[ 
\int_{0}^{T_i}{ \alpha P_c \, e^{(\rho - r)(t + \tau + \sum_{j=1}^{i-1}{T_j})}  v'(\tau) d\tau }   \\ 
&& \qquad
+ \, e^{-r \left(t + \sum_{j=1}^{i}{T_i}\right)} \big[(P_f - \alpha(1-\beta)e^{\rho \left(t + \sum_{j=1}^{i}{T_j} \right) }P_c) \, v(T_i) -C \big]   \bigg] , 
\nonumber
\end{eqnarray}
where
\begin{itemize}
	\item $T_i$ is the length of the i$^{th}$ rotation
	\item $P_c$ is the carbon price at time $t=0$ (\$/tC)
	\item $P_f$ is the price of timber (\$/m$^3$)
	\item $r$ is the real discount rate applied by the forest owner
	\item $\rho$ is the annual real growth rate of the carbon price
	\item $v(\tau)$ is the stem volume at age $\tau$ (m$^3$/ha)
	\item $\alpha$ is a conversion factor between stem volume and total carbon mass (tC/m$^3$)
	\item $\beta$ is the share of carbon stored permanently after fellings
	\item C is the forest regeneration cost (\$/ha).
\end{itemize}

The first term in \eqref{eq:maxproblem} represents the carbon credits accrued from forest growth during the rotation. The increase in forest carbon stock is given by $\alpha v'(\tau)$, valued with a carbon price which increases with the rate $\rho$.
The second term sums over timber revenues with price $P_f$, carbon cost and regeneration costs $C$ at the end of rotation when the forest volume is $v(T_i)$. Both terms are summed over different rotations $i$, and discounted with rate $r$ to time $t=0$.
In order to ensure the present value remains finite, it is required that $r > \rho$, i.e. that the discounting applied by the forest owner is stronger than the growth rate of the carbon price.

The optimal value of the objective function in problem \eqref{eq:maxproblem} -- defined at time $t$ -- can be denoted with a value function $V(t)$. Should the forest's value be based solely on the net present value of revenues and costs from timber, carbon and regeneration; $V(t)$ represents the bare land value at time $t$, discounted to time zero. Using the value function, the problem can be written in a recursive form, where the value of rotations beyond the first is given by $V(t+T_1)$, i.e. the optimal value of the problem \eqref{eq:maxproblem} defined at time $t + T_1$. The recursive formulation is
\begin{eqnarray}
\label{eq:recursiveproblem}
\emph{max}_{\,T_1} 
&& e^{-r t} \Bigg( \alpha P_c e^{\rho t} \int_{0}^{T_1}{ e^{(\rho - r) \tau}  v'(\tau) d\tau }   \\ 
&& \underbrace{
+ \, e^{-r T_1} \left( (P_f - \alpha(1-\beta)e^{\rho (t + T_1)}P_c) v(T_1) -C \right) \Bigg) + V(t + T_1)
}_{:= f(t,T_1)}, 
\nonumber
\end{eqnarray}
where the value function $V(t+T_1)$ returns the optimal value from subsequent rotations, but which depends on the first rotation's length $T_1$.
In this recursive formulation, let us denote the objective function of \eqref{eq:recursiveproblem} with $f(t,T_1)$.

The first-order condition of \eqref{eq:recursiveproblem} for an optimal rotation age $T_1^*$ is
\begin{eqnarray}
\label{eq:FOC}
&&e^{-r (t+T_1^*)} \big( (\alpha  \beta  P_c e^{\rho (t+T_1^*) } + P_f) v'(T_1^*) \\
&&+ (\alpha (1-\beta )(r-\rho )P_c e^{\rho (t+T_1^*)} - r P_f )v(T_1^*) +r C \big) 
\nonumber \\
&&+ \: V'(t+T_1^*)  = 0.
\nonumber 
\end{eqnarray}
This equation does not yet allow the solving of the optimal rotation time $T_1^*$, because it involves $V'(t+T_1^*)$, the derivative of the unknown value function at time $t + T_1^*$. However, an expression for $V'(t)$ can be formulated by using the envelope theorem to the objective function $f(t,T_1)$ at $T_1^*$:
\begin{eqnarray}
\label{eq:envtheorem1}
\frac{dV(t)}{dt} &=& \left. \frac{\partial f(t,T_1^*)}{\partial t} \right|_{T_1^* = argmax_{T_1}\,f(t,T_1)} \\
&=& \alpha P_c (\rho -r) e^{(\rho-r )t}\left(\int_0^{T_1^*} e^{(\rho-r )\tau } v'(\tau)d\tau \right) 
\nonumber \\
&& \, 
- (\alpha P_c (1-\beta)(\rho-r) e^{(\rho-r)(t+T_1^*)} + r P_f e^{-r(t+T_1^*)})v(T_1^*)
\nonumber \\
&& \,
+e^{-r(t+T_1^*)} r C + V'(t+T_1^*).
\nonumber
\end{eqnarray}
The expression for $V'(t)$ in \eqref{eq:envtheorem1} involves the optimal rotation age $T_1^*$, which therefore has to satisfy also the first-order conditions for the optimization problem defined at time $t$. Hence, one can solve $V'(t+T_1^*)$ from \eqref{eq:FOC} and insert this into \eqref{eq:envtheorem1}. This yields
\begin{eqnarray}
\label{eq:envtheorem2}
V'(t) &=& \alpha P_c (\rho -r) e^{(\rho-r )t}\left(\int_0^{T_1^*} e^{(\rho-r )\tau } v'(\tau )d\tau \right) \\
&& \,
- e^{-r (t+T_1^*)}\left(e^{\rho (t+T_1^*) } \alpha  \beta  P_c+P_f\right) v'(T_1^*).
\nonumber
\end{eqnarray}

A final form for the first-order condition is achieved by setting $t$ = 0, using \eqref{eq:envtheorem2} to write an expression for $V'(t+T_1^*)$, and by inserting this into \eqref{eq:FOC}. Now $V'(T_1^*)$ involves also the length of the second optimal rotation $T_2^*$, because $V'(T_1^*)$ associates with the problem defined at time $t$ = $T_1^*$. The first-order condition is then: 
\begin{eqnarray}
\label{eq:FOC_final}
&&
(e^{\rho  T_1^*} \alpha  \beta  P_c+P_f) v'(T_1^*) 
\\ && \nonumber
+ (e^{\rho  T_1^*} \alpha  (1-\beta ) (r-\rho ) P_c - r P_f) v(T_1^*)
\\ && \nonumber
+ e^{\rho T_1^*} P_c\alpha  (\rho -r) (\int _0^{T_2^*}e^{(\rho-r ) \tau } v'(\tau)d\tau ) 
\\ && \nonumber
- e^{-r T_2^*} (e^{\rho  (T_1^*+T_2^*)} \alpha  \beta  P_c+P_f) v'(T_2^*) + r C = 0
\end{eqnarray}

Equation \eqref{eq:FOC_final} involves two unknown variables -- the optimal rotation lengths $T_1^*$ and $T_2^*$ -- in a single equation, and therefore cannot be used alone to solve either of the variables. However, the first order condition can be written for multiple pairs of subsequent rotations -- for $T_1^*$ and $T_2^*$, $T_2^*$ and $T_3^*$, $T_3^*$ and $T_4^*$ and so forth -- in the form of equation \eqref{eq:FOC_final}, all being optimal solutions to the initial problem \eqref{eq:maxproblem}. 
A group of $n$ such equations contains $n+1$ unknown variables, the optimal rotation ages up to $T_{n+1}^*$. 
When this group of equations is solved numerically, $T_{n+1}^*$ can be fixed to some selected value. Although this value is not optimal, due to discounting effects this has only limited impact on $T_1^*$ when $n$ and the difference ($r$ - $\rho$) are sufficiently large, allowing the use of equation \eqref{eq:FOC_final} in approximating the optimal solution of $T_1^*$.

The first order condition \eqref{eq:FOC_final} is a generalization to both the Faustmann and \citet{vanKooten1995} formulae, and reduces to the associated optimality conditions by the insertion of appropriate parameters.
A constant carbon price implies $\rho = 0$. In such a case, the problem setting remains constant over time, and the optimal lengths of consecutive rotations are equal. Therefore, by inserting $\rho = 0$, $T_2^*$ = $T_1^*$, and $C$ = 0 equation \eqref{eq:FOC_final} simplifies upon some algebraic manipulation to the first-order condition of \citet{vanKooten1995}. 
Also, setting $P_c = 0$ eliminates the carbon pricing completely; and again with $T_2^* = T_1^*$ and $C$ = 0, the equation \eqref{eq:FOC_final} simplifies to the first-order condition of the Faustmann rotation.

\section{Numerical examples}
\label{sec:numerical}

Numerical example calculations are presented here with the aim of illustrating the problem setting: how increasing carbon pricing might affect optimal forest rotations, and therefore how forest economics could provide a contribution to economically efficient climate change mitigation.
The growth curves and parameters from \citet{vanKooten1995} are used in order to allow direct comparison with their results.

Forest growth is represented with an idealized function of the form $v(t) = k t^a e^{bt}$.
\citet{vanKooten1995} provided the parameters associated with this functional form for coastal forest in British Columbia and boreal black spruce in Alberta, presented in Table \ref{tbl:params}, including also the parameter $\alpha$ which represents the mass of carbon per volume of timber. 
These functions represent forest volume growth which first gradually increases and then gradually declines. The maximum sustainable yield (MSY) is reached with rotation ages of 90 and 192, and volume reaches its maximum after 122 and 300 years, respectively for the coastal and boreal forest. 
Given that these growth curves are rather generic, future research should further analyse the optimal forest management strategies under some assumed carbon pricing using data from actual forest stands.

\begin{table}[!htb]%
\caption{Growth function parameters $k$, $a$ and $b$, and the carbon content factors $\alpha$ for coastal forest and boreal black spruce, from \citet{vanKooten1995}.}
\label{tbl:params}
\begin{center}
\begin{tabular}{lcccc}
               & $k$      & $a$    & $b$       &  $\alpha$ \\
Coastal forest & 0.000573 & 3.7819 & -0.030965 &  0.1824   \\
Boreal forest  & 0.000759 & 2.7655 & -0.009205 &  0.2030
\end{tabular}
\end{center}%
\end{table}

In these illustrative calculations, the price of timber $P_f$ is set to 50 \$/m$^3$, the discount rate $r$ to 5\%, and the regeneration costs $C$ to zero.
With this parametrization, the Faustmann rotation age is 43 years for the coastal forest and 42 years for the boreal forest.

Regarding carbon pricing, it is necessary to consider a wide range of possible values, because no definitive real-world observations exist for these parameters. The estimates for the social cost of carbon are dispersed \cite[see e.g. ][]{Tol2009}, and prices in existing carbon markets have been very volatile. 
Over time, the price should perhaps grow with a rate close to the marginal productivity of capital if mitigation is carried out in a cost-efficient manner \citep{Ekholm2014}. 
On the other hand, the carbon price growth rate $\rho$ needs to be sufficiently lower than the discount rate $r$, both to keep the objective function of the optimization problem bounded and the approximation of the finite-time problem accurate.
Based on these considerations, the optimal rotation ages are calculated for initial carbon prices $P_c$ up to 150 \$/tC (approximately 41 \$/tCO$_2$) and for growth rates $\rho$ up to 3\%.

For $\beta$ -- the fraction of carbon that is stored permanently after harvest -- the extremes $\beta$ = 0 and $\beta$ = 1 are examined separately. 
This is obviously a simplification of how carbon is stored or released at harvest. Some of the tree biomass is transferred to soil carbon stocks, and only some part can be stored in long-term wood products. Moreover, both soil carbon and wood product stocks gradually decay and release carbon to the atmosphere, but this dynamic aspect is not taken into account in the setting considered here. Therefore the realistic values of $\beta$ are likely to be somewhere between the two extreme values.

The optimal sequence of rotation ages was approximated with the five first rotations, but the results focus on the length of the first rotation. Given that the Faustmann rotation ages for both forest types are over 40 years and that the carbon pricing increases rotation ages, these five rotations span already over 200 years -- at the minimum -- for the growth curves considered here. Situations taking place after such a timeframe are unlikely to have a discernible impact on the optimal length of the first rotations, enabling to use this truncated chain of forest rotations as a reasonable approximation to the original, infinite-horizon problem.

\begin{figure}[!htb]
\begin{center}
\includegraphics[width=0.9\textwidth]{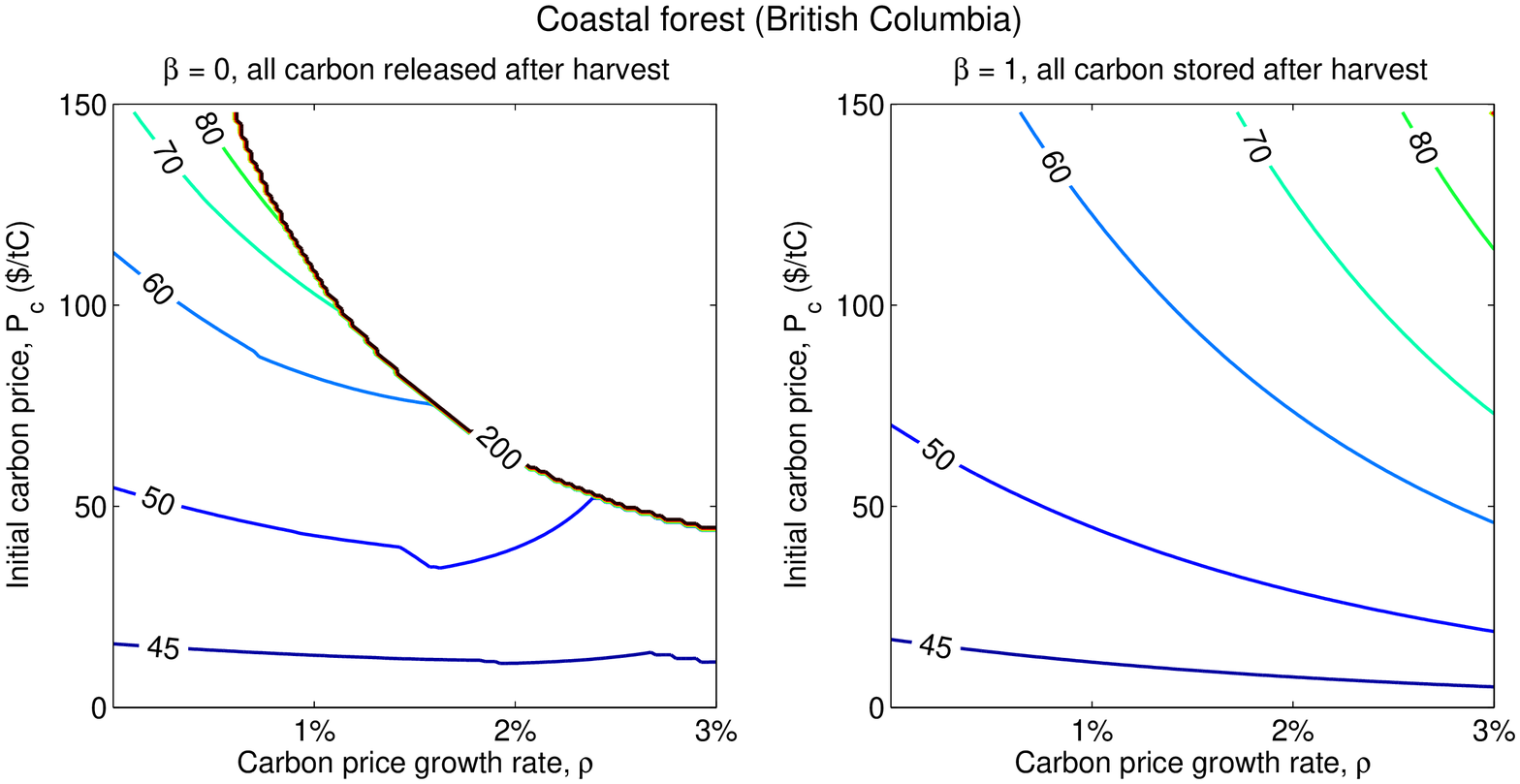}
\\
\vspace{5mm}
\includegraphics[width=0.9\textwidth]{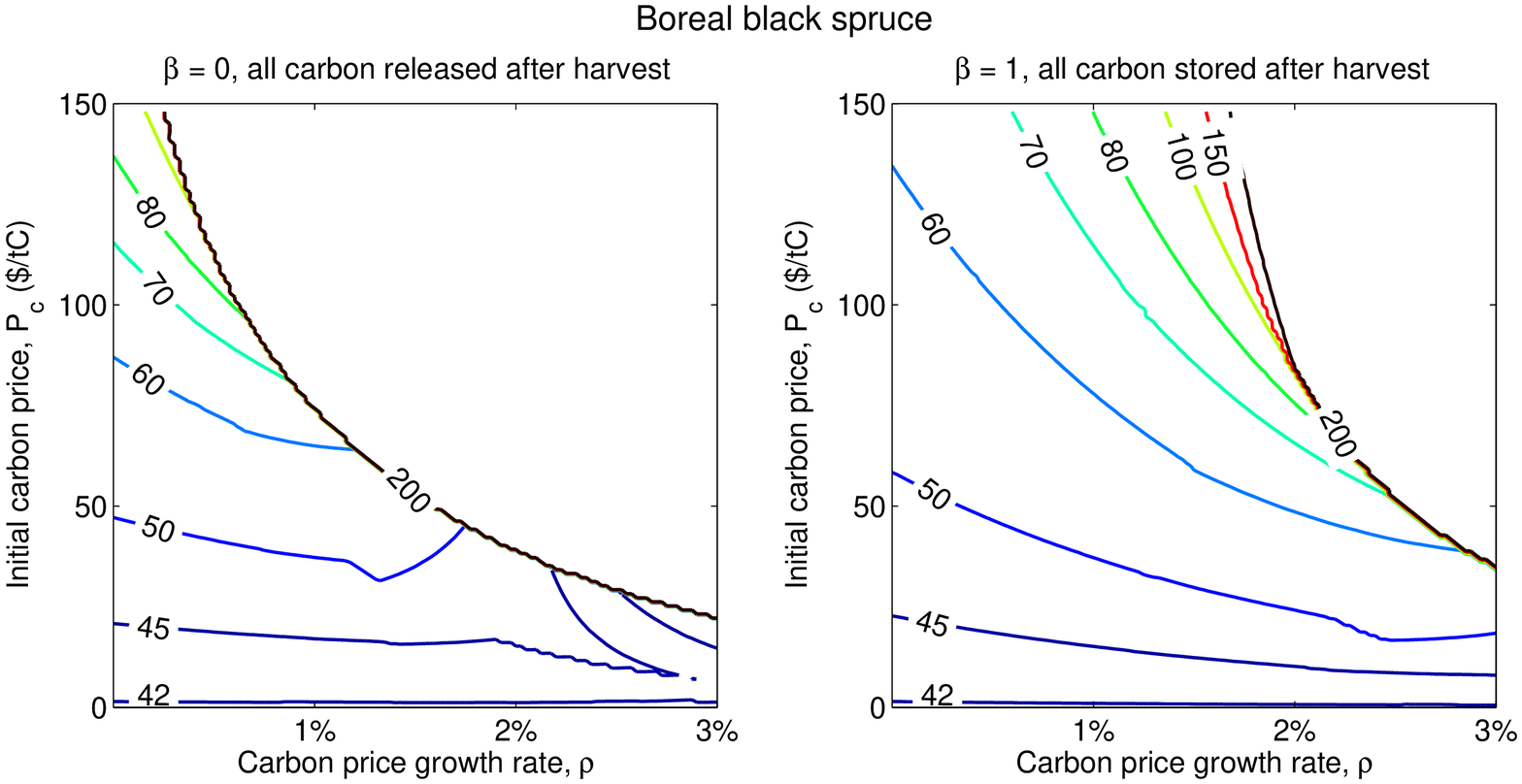}
\end{center}
\caption{The optimal harvest age of the first rotation, when starting from bare land, for different levels of current carbon price $P_c$ and its future growth rate $\rho$. Left and right columns denote respectively cases where all carbon is either released after harvest ($\beta$ = 0), or stored permanently after harvest ($\beta$ = 1).}
\label{fig:OptimalRotation}
\end{figure}

The length of the first rotation in the optimal solution is presented in Figure \ref{fig:OptimalRotation} for the two growth curves, the range of considered values of $P_c$ and $\rho$, and separately for cases with $\beta$ = 0 and $\beta$ = 1.
The bottom of the figures has $P_c$ = 0, and hence corresponds to the Faustmann rotation. Similarly, the left edge has $\rho$ = 0 and corresponds to the rotation lengths presented by \citet{vanKooten1995}.

The general result from all calculated cases is that
$P_c$ and $\rho$ both primarily increase the optimal rotation age. However, this effect is not entirely monotonous, which is visible through the minor ''ripples'', particularly with the black spruce in the case of $\beta$ = 0.
Moreover, the combined effect of $P_c$ and $\rho$ lengthens the rotations further, which is seen from the isocurves of optimal rotation ages being predominantly convex in the $P_c$-$\rho$ plane.

In the $\beta$ = 0 cases, where the harvested carbon is released to the atmosphere, the optimal rotations lengthen to over 200 years with many considered combinations of the $P_c$ and $\rho$ parameters. Rotations lengths over 200 years are not plotted in \ref{fig:OptimalRotation}, as differences between such solutions were considered irrelevant, and the growth curves $v(t)$ might be unreliable to represent actual forest growth beyond multiple centuries.
Therefore, if the carbon price $P_c$ at the time of planting and its future growth rate $\rho$ are sufficiently high, the planted forest would not be harvested for energy or other short-term use for multiple centuries. Despite this, the forest owner receives revenues for the planted forest from the carbon crediting of forest growth.

With $\beta$ = 1, the forest carbon is stored after harvest, and the forest owner is not taxed at harvest as no carbon is released to the atmosphere. In this case, the carbon pricing leads to more moderate increases in the rotation ages for the coastal forest, even with high values of $P_c$ and $\rho$. For the black spruce forest, the rotations are lengthened to over 200 years also with $\beta$ = 1 due to the slow growth of black spruce. As the timber stock slowly accumulates, the carbon price grows if $\rho$ > 0, gradually outweighing the timber price. The growth of black spruce is slow for the first 50 years, but remains considerable even up to the age of 200. Hence harvesting would delay the carbon revenues from forest growth -- which are higher than possible timber revenues -- making harvesting uneconomical.

It is worth to note that these optimal rotation ages generalize easily also to other values of timber price $P_f$. 
Given that here the regeneration costs $C$ were set to zero, the only prices present are for timber and carbon, for which the objective function in \eqref{eq:maxproblem} is linear. Due to this linearity, $P_f$ can be set as a numéraire, and multiplying both prices with a positive constant does not affect the optimal solution. Hence, the optimal rotation ages corresponding to any other timber price can be gained by simply scaling the carbon prices on the y-axis in Figure \ref{fig:OptimalRotation}.

\begin{figure}[!htb]
\begin{center}
\includegraphics[width=0.9\textwidth]{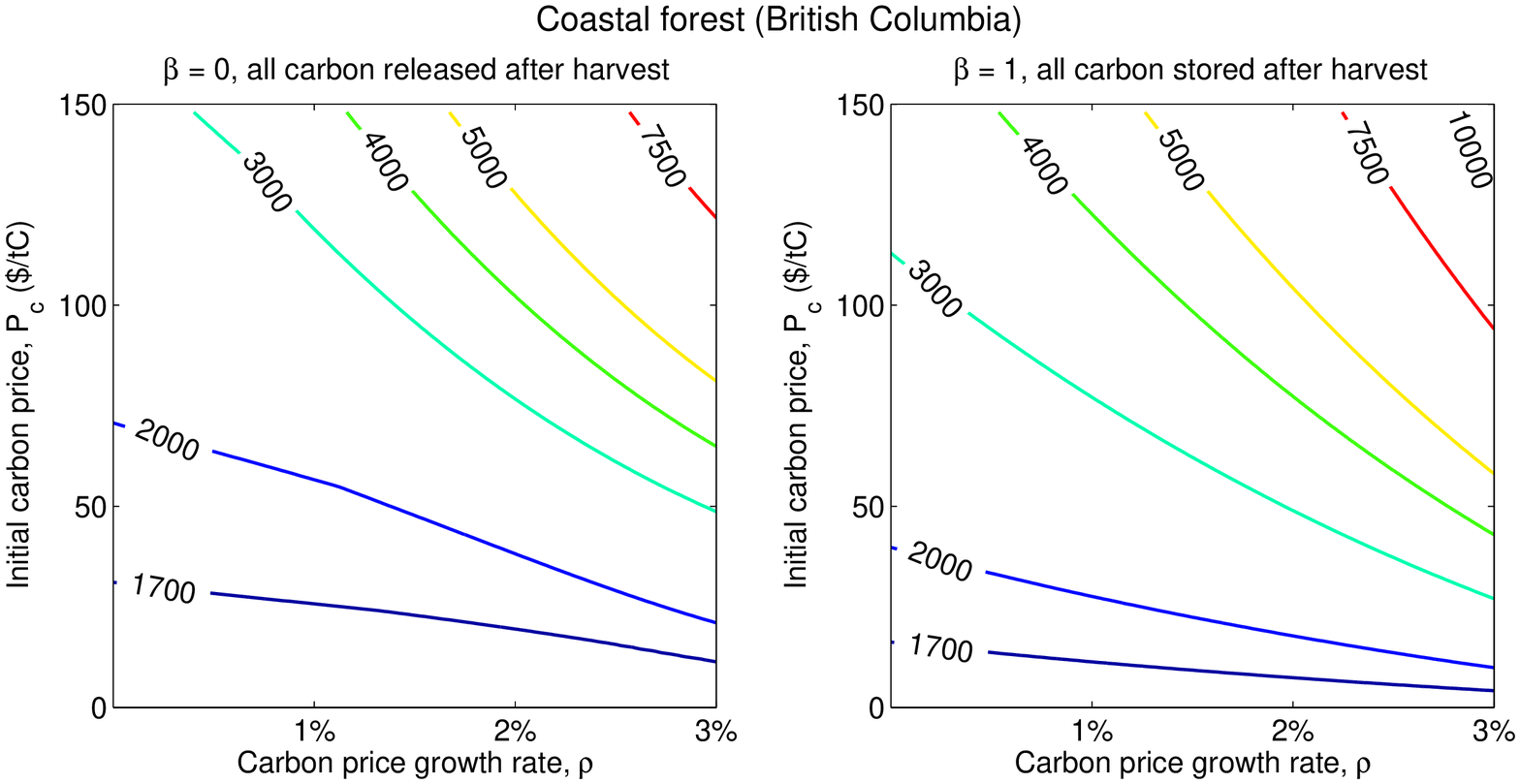}
\\
\vspace{5mm}
\includegraphics[width=0.9\textwidth]{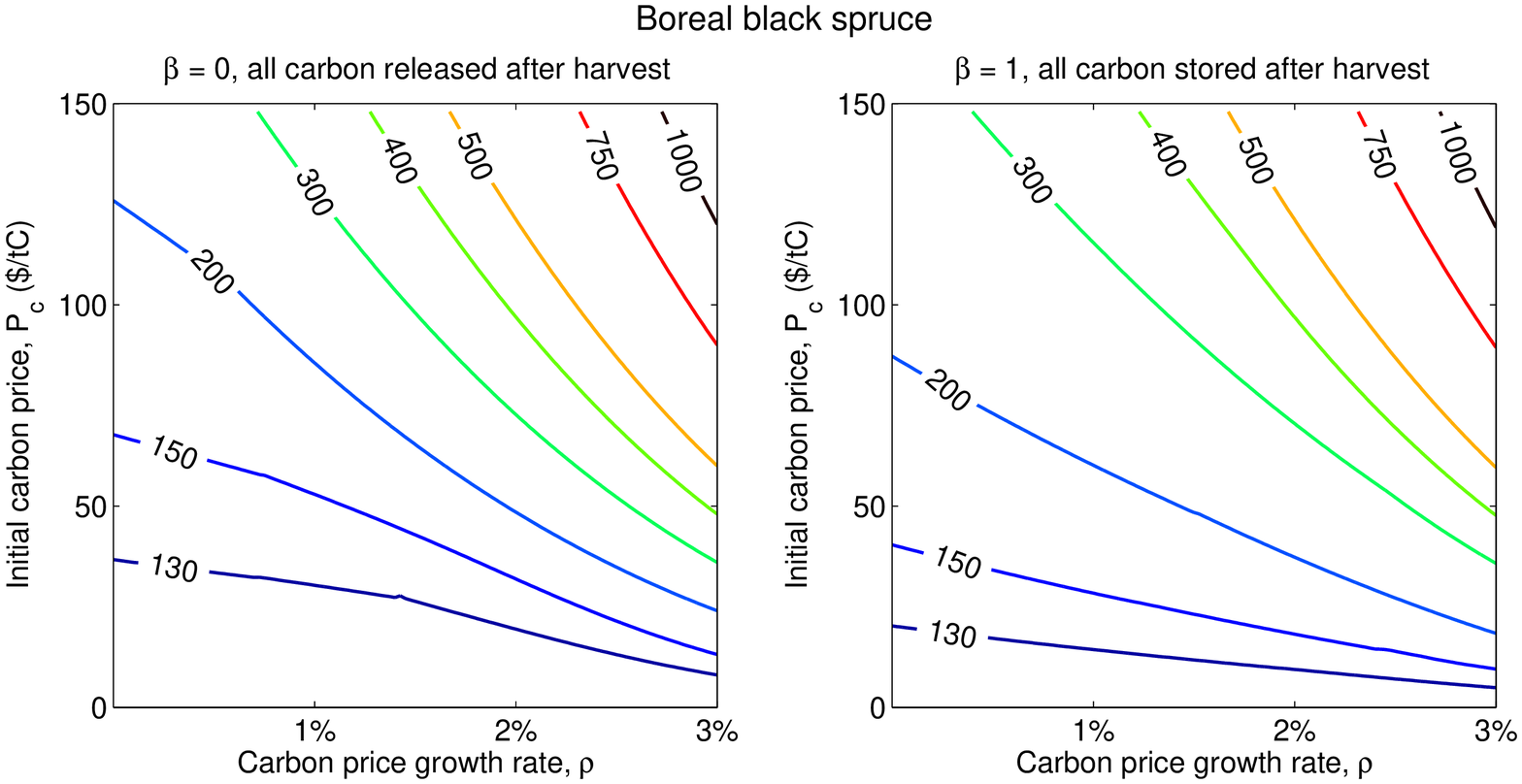}
\end{center}
\caption{The bare land value for different levels of current carbon price $P_c$ and its future growth rate $\rho$. Left and right columns denote respectively cases where all carbon is either released after harvest ($\beta$ = 0), or stored permanently after harvest ($\beta$ = 1).}
\label{fig:OptimalNPV}
\end{figure}

The net present values of bare land corresponding to the optimal sequence of rotations are presented in Figure \ref{fig:OptimalNPV}. 
Carbon pricing increases considerably the revenues of the forest owner from the Faustmann rotation case with $P_c$ = 0, even by an order of magnitude for some of the considered values of $P_c$ and $\rho$. 
For the shorter rotations, the bare land value is higher with $\beta$ = 1 than with $\beta$ = 0, because in this case no carbon tax needs to be paid upon harvest.
This difference illustrates that the carbon pricing mechanism credits the forest owner for permanent storage of carbon
; for temporary storage that lasts only until the next harvest of the forest the revenues can be notably lower.
For the longer rotation lenghts, particularly for those over 200 years, no harvest takes place in a reasonable timeframe and $\beta$ has no discernible effect on the bare land value.

Hence, although that the carbon pricing increases the optimal rotation ages -- or creates outright an incentive to never harvest -- the forest owner's revenues and the bare land value are also increased. Moreover, if the revenues accrue mainly or wholly from carbon pricing, the temporal distribution of revenues is more even than the revenues from timber sales, which occur only at intervals of several decades.

Last, let us look at the problem from a slightly different angle.
The problem setting of equation \eqref{eq:maxproblem} addresses the optimal rotation age for a forest plot that is currently bare land, given that the carbon price is currently at $P_c$ and increases at rate $\rho$.
If the start time $t$ = 0 represents the present, then the first harvests suggested by the results in Figure \ref{fig:OptimalRotation} take place only after several decades or over a century from the present.
Yet, because the optimal rotation ages change over time as the carbon price increases, these results do not answer which forest plots should be harvested now, if carbon price is currently at $P_c$ and increasing in the future with the rate $\rho$.

To address this question, we wish to find the age-classes that are currently at the end of their optimal rotation age. Let $\tau$ be this age, meaning that the forest was planted $\tau$ years ago. 
The optimal rotation age for this particular forest is the result of the problem \eqref{eq:maxproblem} in which carbon pricing starts from a value of $e^{-\rho \tau}P_c$.\footnote{It is not relevant for the optimal solution whether the carbon pricing has actually been in effect or not for the time $\tau$, because the possibly forgone revenues from carbon pricing are comparable to sunk costs, and do not affect optimal decisions into the future.}
Let $T^*(P_c, \rho)$ denote the optimal length of the first rotation -- starting from bare land -- when the carbon pricing at $t$ = 0 is defined by ($P_c$, $\rho$).
The currently optimal harvest age $\tau$ can be solved by using these solutions $T^*(P_c, \rho)$, presented in Figure \ref{fig:OptimalRotation}, by finding the age $\tau$ that satisfies
\begin{equation}
\tau = T^*(e^{-\rho \tau} P_c, \rho),
\end{equation}
where $P_c$ denotes the current carbon price. If a solution to this equation cannot be found due to the jump in optimal rotation ages, there are no age-classes currently at the optimal harvest age.

\begin{figure}[!htb]
\begin{center}
\includegraphics[width=0.9\textwidth]{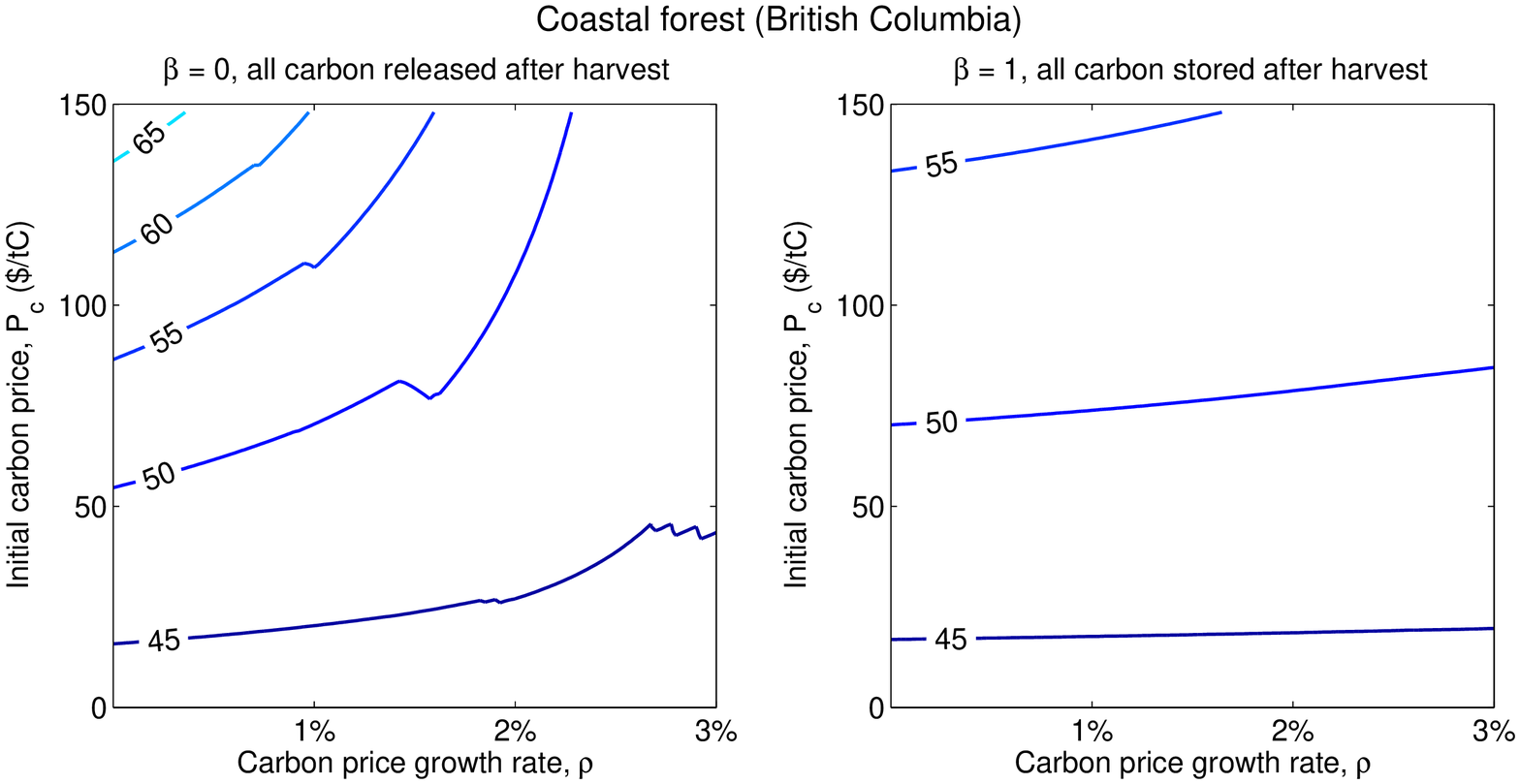}
\\
\vspace{5mm}
\includegraphics[width=0.9\textwidth]{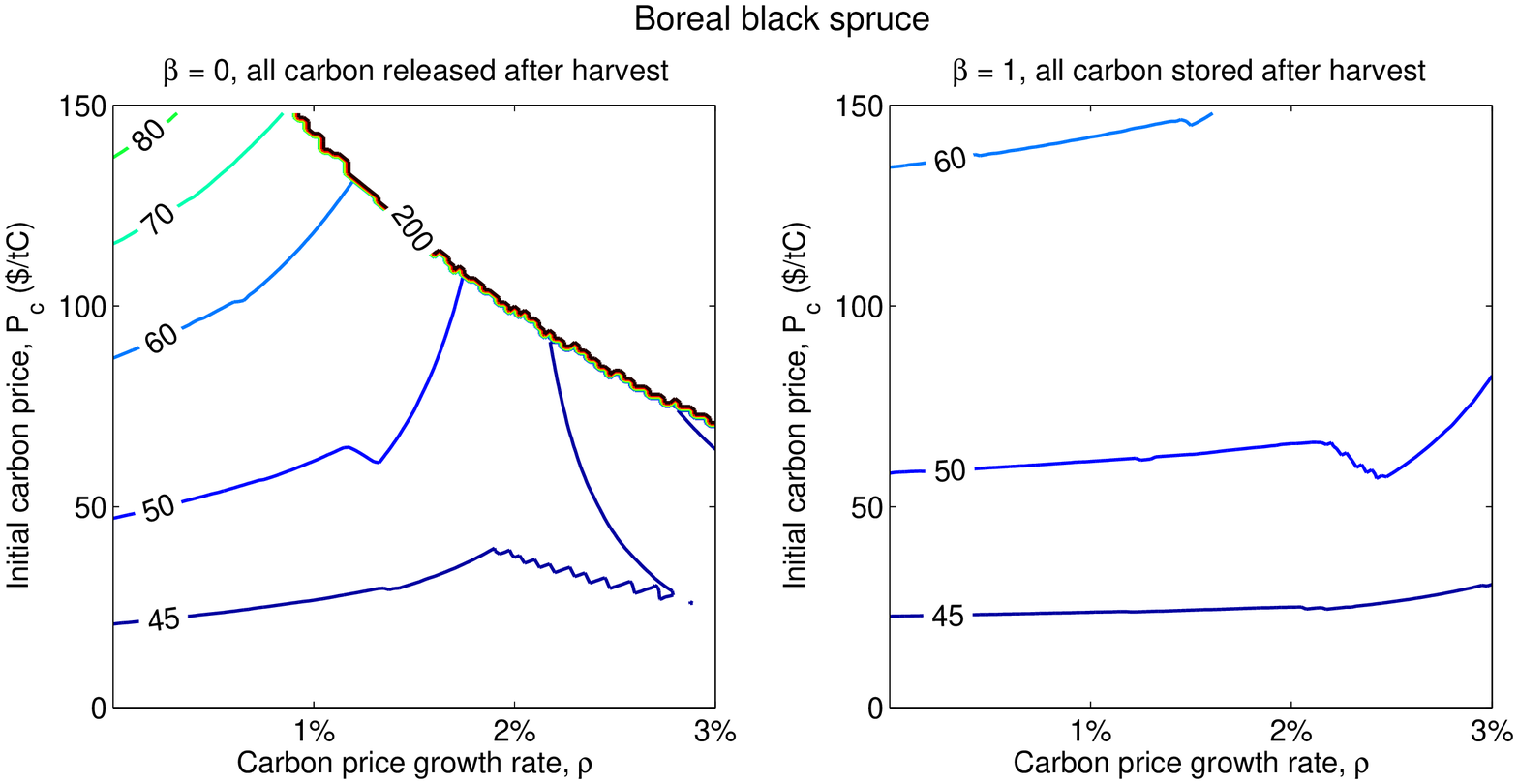}
\end{center}
\caption{The forest age-classes that are currently at the end of their optimal rotation age, for different levels of current carbon price $P_c$ and its future growth rate $\rho$. Left and right columns denote respectively cases where all carbon is either released after harvest ($\beta$ = 0), or stored permanently after harvest ($\beta$ = 1).}
\label{fig:TCurrent}
\end{figure}

These currently optimal harvest ages are presented in Figure \ref{fig:TCurrent}, again for different levels of the current carbon price $P_c$ and its future growth rate $\rho$.
This is essentially Figure \ref{fig:OptimalRotation} with the $P_c$ axis remapped separately for each associated value of $\rho$.
If $P_c$ = 0 or $\rho$ = 0, the optimal rotation age does not change over time, and consequently Figures \ref{fig:OptimalRotation} and \ref{fig:TCurrent} are identical for these cases.

For $P_c$ > 0 and $\rho$ > 0, Figure \ref{fig:TCurrent} is distinctively different from Figure \ref{fig:OptimalRotation}. Higher values of $\rho$ generally decrease -- instead of increasing -- the optimal harvest age. As an implication from this, the impact of carbon pricing on the optimal harvest age is far more moderate than in Figure \ref{fig:OptimalRotation}, 
with the optimal rotation ages exceeding 200 years only with black spruce when $P_c$ and $\rho$ are relatively high and $\beta$ = 0.

Although this effect might seem counter-intuitive, there is an evident rationale behind.
Let us think of a forest plot in the age $\tau$, with $\rho$ > 0.
Because the carbon price is increasing, delaying the harvest would result in that the whole carbon content of the forest is taxed with a higher price at harvest;
and in some instances this can outweigh the other possible net benefits from delaying the harvest.
This effect is not present if $\rho$ = 0, and other thing being equal, the increase in carbon price can create an incentive for a rotation age shorter than in the $\rho$ = 0 case.

Therefore, the introduction of an increasing carbon pricing would lengthen the optimal rotation age of existing forest stands only moderately. For the subsequent rotation, however, the problem starts from bare land and the optimal rotation age corresponds to the results presented in Figure \ref{fig:OptimalRotation}, and hence can be lengthened extensively from the current rotation's optimal length due to the increase in the carbon price during the forest's growth.

Combining these two observations yields an interesting outcome for the higher values of $P_c$ and $\rho$: although harvesting wood for energy or other short-term use might be currently economical, the rising carbon price can render it uneconomical for the subsequent rotations.
Given that currently optimal harvest ages are generally lower for higher values of $\rho$ in Figure \ref{fig:TCurrent},
a higher growth rate for the carbon price would temporally delay the sequestration of carbon to the forest stock, something that has been also reported for the conversion of farmland to a forest carbon storage \citep{vantVeld2005}. Again, for the subsequent rotation a higher value of $\rho$ would generally provide incentives for lengthened rotations, and therefore increased carbon sequestration to the forest.

It is worth to note here, however, that the calculation assumed that all carbon flows are uniformly priced; 
particularly that the whole stock of released carbon is taxed, irrespective whether it has earlier accrued the credits from growth or not.
Enrolling an existing stand in such a carbon pricing program can involve a negative change in the forest's value, due to the large carbon payments due at the possible harvest; while for bare land and young stands the value could potentially increase significantly, as suggested by Figure \ref{fig:OptimalNPV}.

In the context of this paper, land is valued -- in part -- by its potential to sequester and store atmospheric carbon, which implies that the introduction of carbon pricing can indeed have different effects on the value of bare land and existing forests. Should e.g. only the carbon stock sequestered after a stand has entered the carbon pricing scheme be subject to the carbon tax at harvest, this could potentially change also the currently optimal harvest ages presented in Figure \ref{fig:TCurrent}.

\section{Discussion and conclusions}
\label{sec:conclusions}

This paper has analysed the harvest age of even-aged forest rotations that maximize the net present value of revenues for the forest owner under constant timber prices and an increasing price for carbon, which is generally associated with economically efficient climate change mitigation.
Compared to the constant carbon price case of \citet{vanKooten1995}, an increasing carbon price predominantly increases the length of optimal rotations -- and in some cases significantly -- particularly if the carbon is released to the atmosphere upon harvest.
The combined effect of the initial price and the growth rate strengthens the lengthening further.
In the near-term, however, when considering stands that are already close to the optimal rotation age, a higher growth-rate for the carbon price can also lead to optimal rotation lengths that are shorter than what a constant carbon price would imply.
Therefore, due to discounting and forest growth dynamics, the effect of a growing carbon price on optimal forest rotation length seems to be non-trivial.

The calculation of bare land value assumed that the land-owner can reap the benefits from the carbon pricing in full, and resulted in that the bare land value could increase by almost an order of magnitude with the highest considered values of the initial carbon price and its growth rate. 
This increase in land value was higher in the case that the harvested carbon could be stored permanently, and lower for temporary storage of carbon \citep[see also][]{ Kim2008}.
It is worth to note that in the parametrizations with which harvests did not take place within multiple centuries, the value of bare land was driven solely by the land's potential to store carbon.
However, for existing, older stands the introduction of such carbon pricing could lower the forest's value, because the credits from forest growth would have already been forgone but the harvest would entail a full payment for all released carbon.

The paper started from the premise that all carbon flows should be priced with a uniform price across the economy to achieve an economically efficient mitigation strategy. 
To interpret the numerical examples' results in this context, an efficient mitigation strategy would be associated with increased carbon sequestration by forests through lengthened rotations, and also land being valued by its ability for capturing and storing atmospheric carbon. The latter effect is not limited to current forest land only: a uniform carbon price would similarly provide incentives e.g. for afforestation of non-forest land.

The approach employed here was a somewhat theoretical simplification, and additional bits of realism should be added to the analysis in subsequent research. 
Only part of the carbon stored in the living tree biomass can be harvested, and the biomass stored in soil carbon stock gradually decays to the atmosphere. 
\citet{Hoel2014} found that under constant carbon pricing, the consideration of multiple carbon pools extends the optimal rotation period, and this analysis could be extended to an increasing carbon price, as was assumed here. Similarly, the permanence of harvested wood products -- which was simplified to the parameter $\beta$ in this paper -- should be taken better into account to arrive at a more comprehensive view of forest-related carbon flows.

The long timeframes involved in both climate change mitigation and forest management render the setting susceptible to changing situations.
Timber price was held constant in this paper, and the analysis should be extended to cover a varying price also for timber \citep{Chang1998, Guthrie2009, Susaeta2014}.
Yet, the bases for timber and carbon pricing are fundamentally different. Here it was assumed that the increase in carbon price arises directly from the problem formulation for climate change mitigation; while timber price is determined by demand and supply. How this balance develops in the future is an open question, but some of the lengthening of rotations under increasing carbon pricing might be countered if the willingness to pay for timber would increase over time.
In addition, carbon markets are not established as those for timber, rendering the carbon revenues perhaps more unreliable than those from timber sales.
The carbon price can deviate significantly from the expected path \citep{Ekholm2014}, and the perfect-foresight assumption in the problem formulation is not adequate to address the possible risks involved.

The risk of involuntary and unanticipated release of carbon, e.g. due to fires \citep{Couture2011} or the changing climate \citep{Galik2009}, should also be taken into account. 
Although the voluntary release of carbon was taxed in the problem setting analysed here -- and although an averaged estimate for forest damages could be incorporated already in the assumed growth curve -- an explicit consideration of such events would enable to address the non-permanence features of forest carbon stocks more comprehensively.

Moreover, paying for the carbon flows is not the only way to incentivize the forest carbon sequestration.
Other approaches -- such as carbon rentals and investment subsidies -- have also been analysed in past research \citep{Marland2001, Sohngen2003, Uusivuori2007}. 
The carbon payments can also be based on additionality, which could avoid from paying windfall profits to the forest owner. However, additionality is based on some assumed counterfactual forest management strategy which can be difficult to determine in practice, and might not value land fully by its carbon storage potential.
Also, other forest management options apart from even-aged harvests \citep[see e.g.][]{Stainback2002, Pohjola2007} should be examined also in the setting of increasing carbon prices.

The large impact of carbon pricing on optimal rotation length and land value indicates that forestry can have an important contribution to the climate change mitigation though increased sequestration of carbon, provided that appropriate incentives are in place.
Forests serve numerous functions -- such as the provision of material, energy, habitat and carbon storage -- and introducing a new source of value would inevitably change the balance between these functions.
Balancing the different functions requires intricate evaluation of these different values, which cannot be done in isolation from the surrounding ecosystem, climate, and economic factors.
Such considerations should be incorporated to the problem setting in subsequent research, 
both to better understand the factors affecting optimal forest management,
and also to identify novel ways how forestry could ideally support economic welfare, ecosystems and climate change mitigation.

\section*{Acknowledgements}

The research has been done in the project ECOSUS, funded by the Academy of Finland (decision n:o 257174).

\section*{References}

\bibliographystyle{agsm}
\bibliography{references}

\end{document}